\documentclass[%
 aip,
 amsmath,amssymb,
 reprint,%
]{revtex4-1}

\usepackage{graphicx}
\usepackage{dcolumn}
\usepackage{bm}

\usepackage[utf8]{inputenc}
\usepackage[T1]{fontenc}
\usepackage{mathptmx}
\usepackage{etoolbox}

\draft 

\begin{document}


\title{Continuous-variable quantum key distribution over 28.6 km fiber with an integrated silicon photonic receiver chip} 



\author{Yiming Bian}
\thanks{These two authors contribute equally to this work.}
\affiliation{State Key Laboratory of Information Photonics and Optical Communications, Beijing University of Posts and Telecommunications, Beijing, 100876, China}
\affiliation{School of Electronic Engineering, Beijing University of Posts and Telecommunications, Beijing, 100876, China}

\author{Yan Pan}
\thanks{These two authors contribute equally to this work.}
\affiliation{Science and Technology on Communication Security Laboratory, Institute of Southwestern Communication, Chengdu 610041, China}

\author{Xuesong Xu}
\affiliation{State Key Laboratory of Information Photonics and Optical Communications, Beijing University of Posts and Telecommunications, Beijing, 100876, China}
\affiliation{School of Electronic Engineering, Beijing University of Posts and Telecommunications, Beijing, 100876, China}

\author{Liang Zhao}
\affiliation{State Key Laboratory of Information Photonics and Optical Communications, Beijing University of Posts and Telecommunications, Beijing, 100876, China}
\affiliation{School of Electronic Engineering, Beijing University of Posts and Telecommunications, Beijing, 100876, China}

\author{Yang Li}
\affiliation{Science and Technology on Communication Security Laboratory, Institute of Southwestern Communication, Chengdu 610041, China}

\author{Wei Huang}
\affiliation{Science and Technology on Communication Security Laboratory, Institute of Southwestern Communication, Chengdu 610041, China}

\author{Lei Zhang}
\affiliation{State Key Laboratory of Information Photonics and Optical Communications, Beijing University of Posts and Telecommunications, Beijing, 100876, China}
\affiliation{School of Integrated Circuits, Beijing University of Posts and Telecommunications, Beijing, 100876, China}

\author{Song Yu}
\affiliation{State Key Laboratory of Information Photonics and Optical Communications, Beijing University of Posts and Telecommunications, Beijing, 100876, China}
\affiliation{School of Electronic Engineering, Beijing University of Posts and Telecommunications, Beijing, 100876, China}

\author{Yichen Zhang}
\email[]{zhangyc@bupt.edu.cn}
\affiliation{State Key Laboratory of Information Photonics and Optical Communications, Beijing University of Posts and Telecommunications, Beijing, 100876, China}
\affiliation{School of Electronic Engineering, Beijing University of Posts and Telecommunications, Beijing, 100876, China}

\author{Bingjie Xu}
\email[]{xbjpku@163.com}
\affiliation{Science and Technology on Communication Security Laboratory, Institute of Southwestern Communication, Chengdu 610041, China}


\date{\today}

\begin{abstract}
    Quantum key distribution, which ensures information-theoretically secret key generation, is currently advancing through photonic integration to achieve high performance, cost reduction and compact size, thereby facilitating the large-scale deployment. 
    Continuous-variable quantum key distribution is an attractive approach for photonic integrations due to its compatibility with off-the-shelf optical communication devices. However, its chip-based systems have encountered significant limitations primarily related to the shot-noise-limited receiver design, which demands low noise, wide bandwidth, high clearance and well stability.
    Here, we report the implementation of a real local oscillator continuous-variable quantum key distribution system with an integrated silicon photonic receiver chip. Thanks to the well-designed chip-based homodyne detectors with a bandwidth up to 1.5 GHz and a clearance up to 7.42 dB, the transmission distance of the system has been extended to 28.6 km, achieving a secret key generation rate of Mbps level. 
    This technological advancement enables the quantum key distribution systems with photonic integrated receivers to achieve the coverage in both access network scenarios and short-distance metropolitan interconnections, paving the way for the development of the next-generation quantum key distribution networks on a large scale.
\end{abstract}

\pacs{}

\maketitle 

Quantum key distribution (QKD) \cite{bennet1984quantum, AdvInQC, PTPQKDRMV2020} allows secret key generation between two remote parties with provable information-theoretically security by leveraging the principles of quantum mechanics. It can be implemented with discrete variable (DV) or continuous variable (CV) \cite{GG02Nature,GaussianQuantumInformation,lam2013continuous}, such as the polarization of single photons or the quadrature of squeezed states, respectively.
Among which, coherent-state CV-QKD \cite{GG02PRL,NSPRL} offers easy implementation with off-the-shelf optical devices working at room temperature \cite{CvExp80km2012,CVMDIYork,CVQKD50km,CvExp202kmPRL,CVNC2022,hajomer2024long}.
Its well compatibility with the existing telecommunication infrastructures allows the utilization of state-of-the-art encoding \cite{DMCVLeverrier,DMCVLinjie} and decoding \cite{SubGbps,Pi2023SubMbps} techniques in modern telecommunications, making it highly accessible for practical deployments \cite{CVReV2023}.

Advances in photonic integrated circuits (PICs) have greatly promoted the developments of QKD systems, which offer the advantages including compactness, cost effectiveness and ease of massive production \cite{wang2020integrated,luo2023recent}.
The footprint of the chip-based optical components can be reduced to several square millimeters, making it suitable to be deployed into the scenarios with requirements on system size, such as the satellites. Integration of the components onto a chip streamlines the production process, which significantly reduces the overall costs of mass productions. 
Among various PIC platforms, silicon photonics has promising prospects. Leveraging the complementary metaloxide-semiconductor (CMOS) fabrication, silicon photonics benefits from the well-established semiconductor industry infrastructure, where the manufacturers can use existing facilities and processes, resulting in lower production costs compared to other photonic materials \cite{soref2006past,lim2013review,siew2021review}.

\begin{figure*}
    \includegraphics[width= 18 cm]{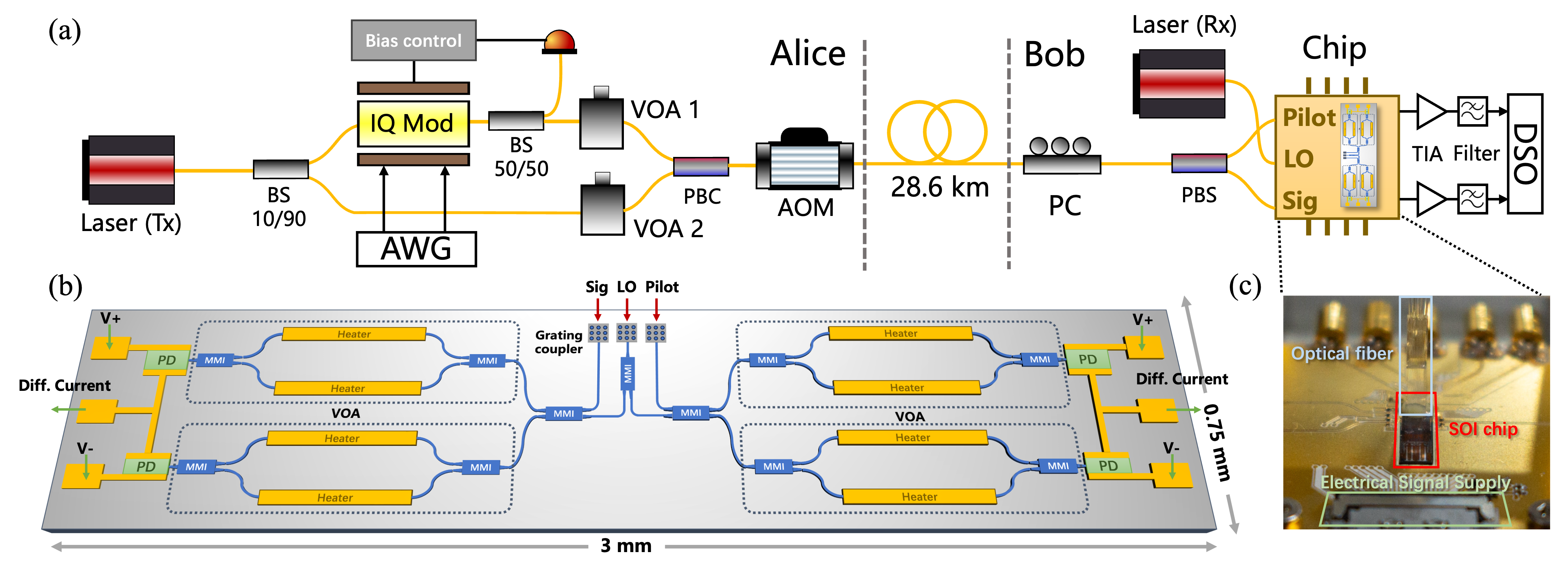}%
    \caption{Optical layout of the CV-QKD system with an integrated silicon photonic receiver. (a) The system set up. IQ Mod: IQ modulator, VOA: variable optical attenuator, AOM: acousto-optic modulator, PBC: polarization beam coupler, PC: polarization controller, PBS: polarization beam splitter, TIA: trans-impedance amplifier, AWG: arbitrary waveform generator, DSO: digital storage oscilloscope. (b) The integrated silicon photonic receiver. PD: photodiode, MMI: multi-mode interferometer. (c) The integrated silicon photonic receiver after packaging. \label{fig:chip}}%
\end{figure*}

The measurement of quantum states plays a crucial role in a QKD system, therefore, its integration holds significant importance. Unfortunately, achieving a high-performance QKD system with silicon photonic receivers remains challenging, which limits the development of the fully integrated systems.
Until now, the CV-QKD system, which benefits from homodyne detection \cite{bruynsteen2021integrated,tasker2021silicon,jia2023silicon}, has been successfully demonstrated to be feasible with a silicon photonic receiver in an in-line local oscillator (LO) scheme \cite{CvExpSOICVQKD2019} and a real LO scheme \cite{pietri2023cv}.
However, the transmission distance and secret key rate are limited to 10 km and 100 kbps level by the unsatisfactory receivers, which significantly reduces the practicality.
Towards better system performance, the noise and fluctuations should be further suppressed, and a silicon photonic receiver that has wide bandwidth, low noise and high stability is necessary.

Here we report a high-performance real LO CV-QKD system with a silicon photonic receiver, which significantly enhances the practicality and effectiveness of chip-based QKD systems. 
The specially designed homodyne detector with a bandwidth of 1.5 GHz and a clearance over 7.42 dB allows the measurement of quantum signals with high repetition frequency up to 1 GHz. 
The feedback control modules on chip and the well designed photonic and electronic amplification circuits contribute to a plat shot noise spectrum and stable detection performance, which enable the suppression of the receiver noise. As a results, the transmission distance of the system has been extended to 28.6 km and the asymptotic secret key rate is over 1.3 Mbps. For the first time, a QKD system with a silicon photonic receiver has achieved a transmission distance over 25 km and a secret key rate at Mbps level, which signifies that the QKD system on chip now has the ability to fully cover the short-range interconnections, and is heading towards the mid-range metropolitan links.

The system set up is shown in Fig. \ref{fig:chip} (a). At Alice's site, coherent light is generated by a continuous-wave laser with a linewidth of 0.1 kHz (NKT Photonic Basik X15), where the wavelength is set to 1550.12 nm. It is then split into a weak signal path and a strong pilot signal path by a 10/90 beam splitter. In the signal path, the light is modulated by an In-phase/quadrature (IQ) modulator (AFR 40 G-IQ) that works at the optical single sideband modulation with carrier suppression by an automatic bias controller. The electrical Gaussian signals that drive the IQ modulator are generated from the two-channel arbitrary waveform generator (AWG) (Keysight M8190A) with the sample rate of 5 GSa/s and a $750$~MHz frequency shift, where the repetition rate of the system is set to 1 GHz. After that, the modulated optical signals are attenuated to a several-photon level by a variable optical attenuator (VOA). With a polarization beam combiner, it is combined with the pilot signals of which the power is controlled by another VOA. In this way, the weak quantum signals and the strong pilot signals are multiplexed in two dimensions, polarization and frequency, to suppress the crosstalk.
Subsequently, an acousto-optic modulator (AOM) with an extinction ratio of 50 dB is employed to control the on-off of the multiplexed optical signals, where the fast switching allows the real-time shot-noise unit (SNU) calibration, which reduces the impact of the shot noise fluctuations on system performance. 

Bob's site consists of the polarization de-multiplexing module, the LO laser and the integrated silicon photonic receiver which is manufactured with the SITRI 180nm process node on a 8-inch (20.3 cm) silicon-on-insulator wafer with 220nm silicon layer \cite{SITRI}.
The key components of the integrated receiver, including the PDs, interference structures and the VOAs, are integrated on a silicon photonic chip with a footprint of $3 \times 0.75 \ mm^2$.
In the polarization de-multiplexing module, the quantum signals and the pilot signals are de-multiplexed by a polarization beam splitter placed after a polarization controller. The LO is generated by an independent laser (NKT Photonic Basik X15) with about 1.5 GHz frequency offset from Alice's laser. Subsequently, the quantum signal, LO and pilot signals are coupled into the chip by three grating couplers with insertion loss of 4 dB. As shown in Fig. \ref{fig:chip} (b), the LO is split by a $1\times 2$ multi-mode interferometer (MMI) and then interfered with quantum and pilot signals in the $2 \times 2$ MMI, respectively. 
After each $2 \times 2$ MMI, two VOAs with a Mach-Zehnder interferometer (MZI) structure are deployed to balance the two arms. The extinction ratio of the VOA is controlled by the voltages applied to the heaters, which adjust the temperature of the waveguide and phase shift based on thermo-optic effect.
The outputs are then detected by two balanced homodyne detectors comprising cascaded photodiodes (PD) with a responsivity of 0.8 A/W, which outputs the differential current that contains the homodyne detection result. 
Each differential current is then sent into a trans-impedance amplifier (TIA) with 40 dB gain, 160 ps rise time and low electronic noise. The voltages of the VOAs are adjusted based on the feedback from TIA to suppress the direct current and reduce the common mode rejection ratio of homodyne detection. The output of the TIA is then sent into a filter circuit to optimize the response in frequency domain. Finally, the output signals of the chip-based receiver are digitalized by a digital storage oscilloscope (DSO) (Keysight DSO S404A) with 5 GSa/s sample rate.

\begin{figure}
    \includegraphics[width= 8 cm]{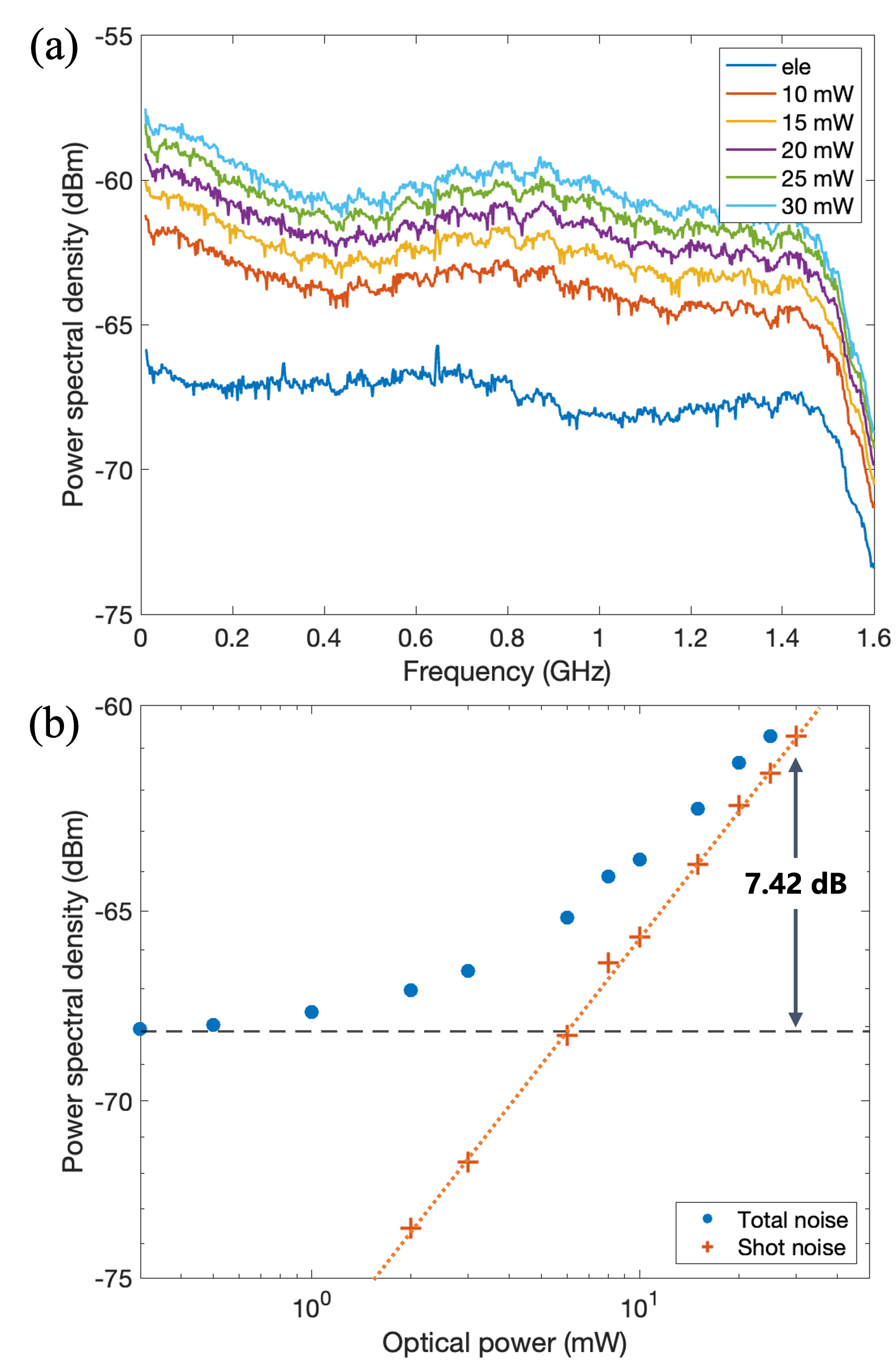}%
    \caption{The noise power spectral density (PSD) of the receiver. (a) The PSD of electronic noise, and total noise with different LO power. (b) The PSD measured at 1 GHz with different LO power. Measurements are performed with a Keysight N9000B CXA Signal Analyzer. The LO power is measured before sent into the receiver chip. \label{fig:Spec}}%
\end{figure}

As shown in Fig. \ref{fig:Spec}, the bandwidth of the chip-based receiver can reach 1.5 GHz, and the clearance of 7.42 dB is achieved.
Fig. \ref{fig:Spec} (a) shows that, the fluctuations of the total noise and electronic noise are around 3 dB and 2.4 dB respectively. Fig. \ref{fig:Spec} (b) reflects the well linearity of the homodyne detector.
These advancements are achieved by the balance control with high accuracy, the consistency of the chip-based optical components, and the low-noise amplification circuits.
The flat spectrums of electronic noise and shot noise result in a better detection performance and stability, which contribute to the suppression of the excess noise at the receiver site. Since the total excess noise can be expressed as $\varepsilon_{tot} = \varepsilon_{A} + \varepsilon_{B} / T_{tot}$, where $\varepsilon_{A}$ and $\varepsilon_{B}$ are the excess noise introduced at the transmitter and receiver site, and $T_{tot} < 1$ is the overall transmittance of the channel. $\varepsilon_{tot}$ is more sensitive to $\varepsilon_{B}$, especially in long-distance and high-loss scenarios, where $T_{tot}$ is extremely low. Therefore the suppression of the excess noise introduced by the receiver is crucial for the extension of the transmission distance. 

\begin{figure}
    \includegraphics[width= 7 cm]{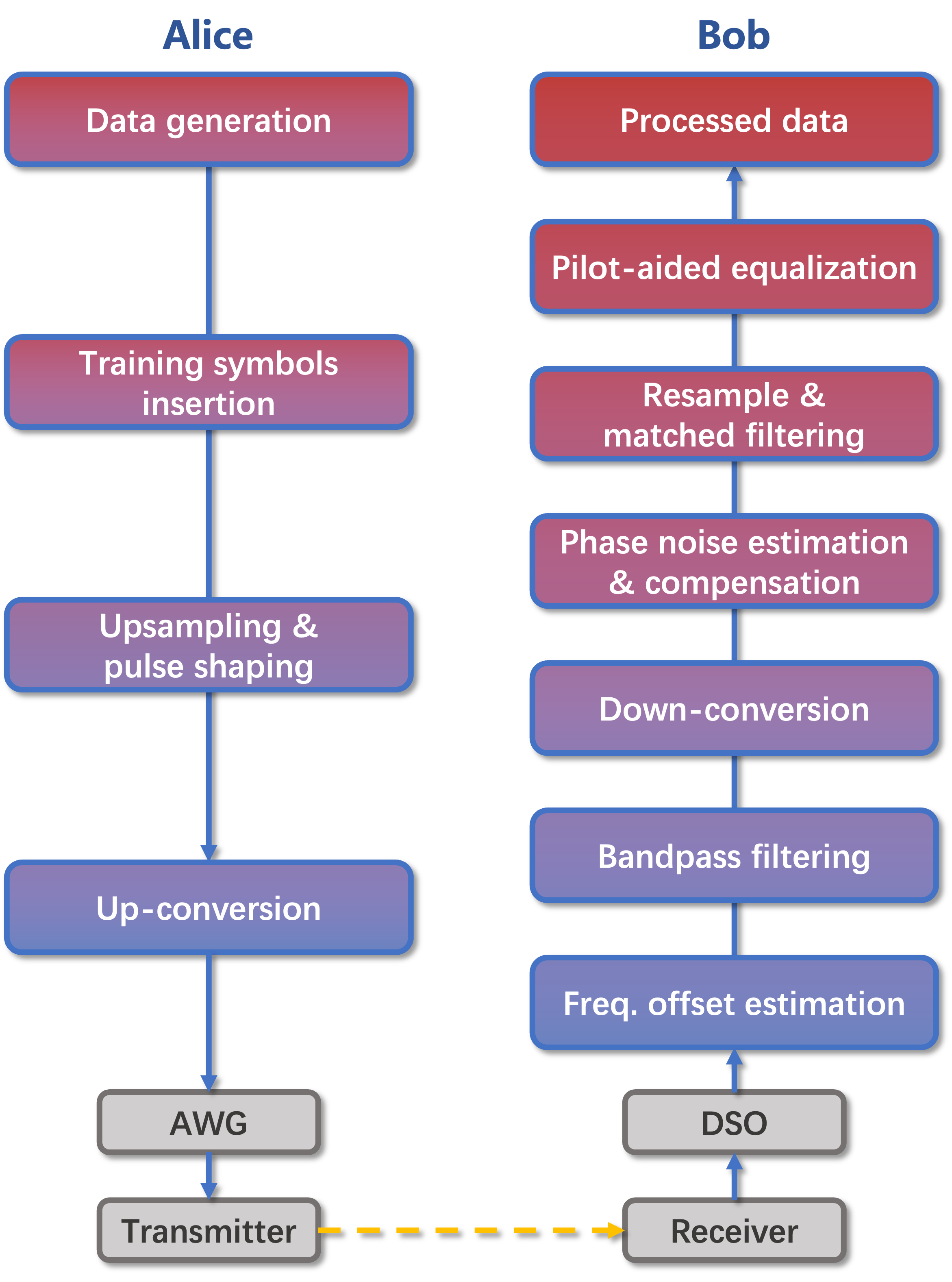}%
    \caption{DSP routine of the system. The roll-off factor is 0.3 for pulse shaping and RRC filter, bandwidth of the bandpass filters for quantum and pilot signals are set at 1.3 GHz and 200 kHz, respectively. AWG: arbitrary waveform generator, DSO: digital storage oscilloscope. \label{fig:DSP}}%
\end{figure}

Digital signal processing (DSP) is of significant importance for a high-speed system. As shown in Fig. \ref{fig:DSP}, Alice's modulation signal is obtained by the following steps: 
1) Gaussian data generation. The random bits produced by the quantum random number generator (QRNG) are converted to uniform distributed decimal random numbers with every 16 bits, and then the Gaussian data can be generated by the Box-Muller transform; 
2) Training symbols insertion. For training the optical channel impairments, a deterministic sequence of QPSK pilot symbols is interleaved in time with the Gaussian data, and the transmitted data can be obtained; 
3) Upsampling and pulse shaping. The transmitted data is upsampled to 5$\times$ oversampling (zero insertion), and a root raised cosine filter with a roll-off factor of 0.3 is used for pulse shaping; 
4) Digital up-conversion. To realize the frequency domain multiplexing of quantum and reference signal, digital up-conversion is performed to the shaped signal. Then, the processed signal is loaded to the arbitrary waveform generator, which is working at 5 GSa/s.
As for Bob, the offline DSP procedure mainly includes 
1) Frequency offset estimation. The frequency offset between the carrier laser at the transmitter and the local laser at the receiver is set to about 1.55 GHz for the intermediate frequency signal detection. Due to the wavelength shift of the lasers, frequency offset estimation should be performed to obtain the accurate center frequency of quantum and reference signals; 
2) Bandpass filtering. To filter the out-of-band noise, a frequency-domain ideal bandpass filter with bandwidth of 1.3 GHz and 200 kHz is used for quantum and reference signals, respectively. 
3) Digital down-conversion. The x and p quadrature of quantum and reference signal are demodulated from the intermediate frequency signal by digital down-conversion; 
4) Phase noise estimation and compensation. Generally, the high-power reference signal has a similar phase noise as the quantum signal. Thus, the phase noise is estimated by the high-power reference signal, and compensated to the quantum signal. 
5) Resample and matched filtering. The phase-compensated signal is resampled to 4$\times$ oversampling, and an RRC filter with a roll-off factor of 0.3 is used to filter the baseband quantum signal. 
6) Pilot-aided equalization. To compensate for the imbalance of x and p quadrature, residual inter-symbol interference, and residual phase noise, a real-valued finite-impulse response (FIR) filter is implemented. 

\begin{figure}
    \includegraphics[width= 9 cm]{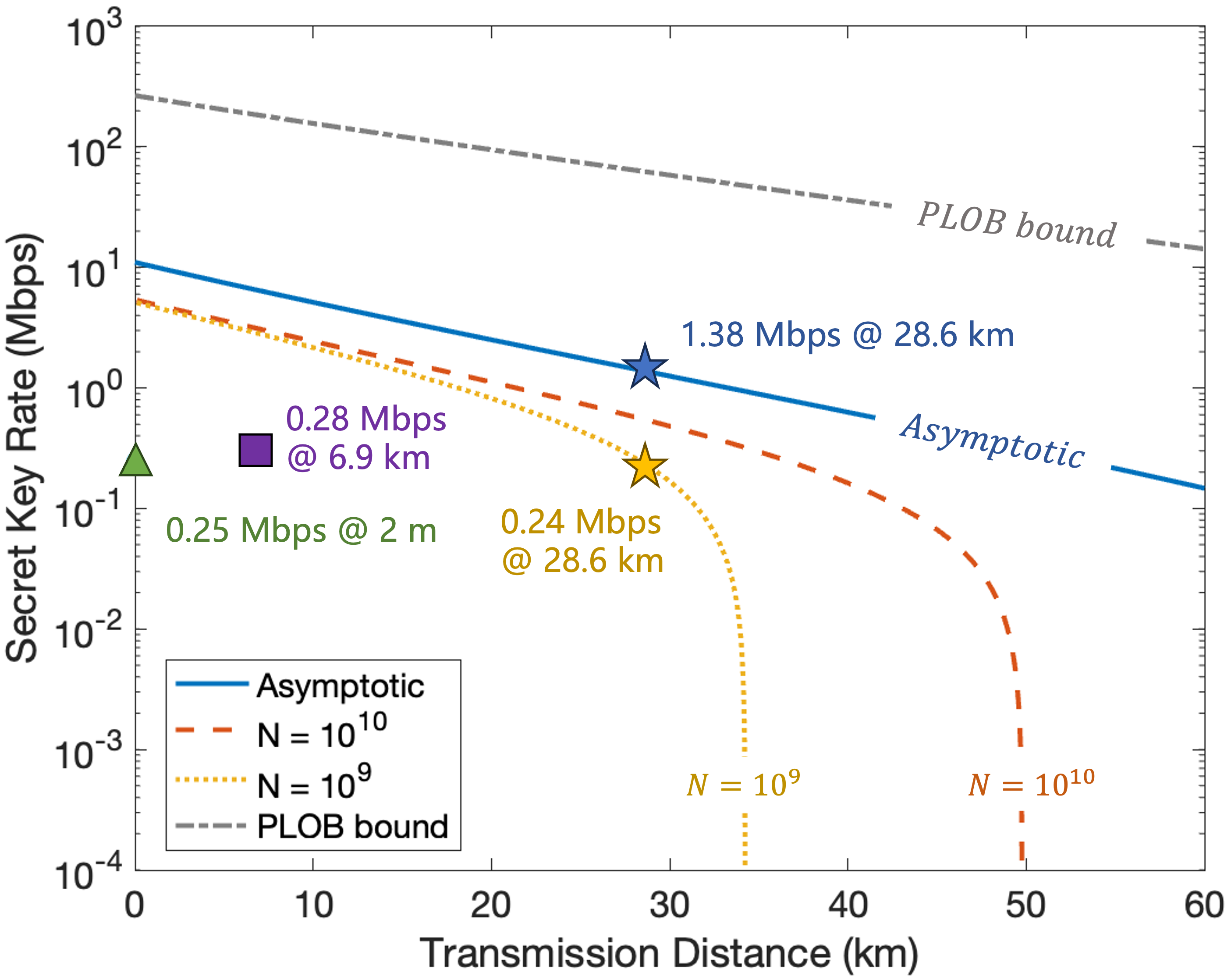}%
    \caption{Experimental secret key rate and simulations. The solid blue, dashed red and dotted yellow curves represent the asymptotic and the finite-size secret key rate with data length of $10^{10}$ and $10^9$. PLOB bound \cite{pirandola2017fundamental} is plotted with respect to the effective repetition rate of 500 MHz and an extra untrusted loss of 5 dB. The stars are the experimental points at 28.6 km, which are compared with the existing experimental results represented by triangle \cite{CvExpSOICVQKD2019} and square points \cite{pietri2023cv}. \label{fig:SKR}}%
\end{figure}

After the above processing, security analysis of the achieved raw data is performed to evaluate the system performance based on
\begin{equation}
    K = f(1-a)[\beta I_{AB}-\chi_{BE}-\Delta(n)].
\end{equation}
Here, $K$ is the secret key rate, $f$ is the repetition frequency, $a$ is the system overhead, $\beta$ is the reconciliation efficiency, $I_{AB}$ is the classical mutual information between legitimate parties, $\chi_{BE}$ is the upper limit of the mutual information between the eavesdropper Eve and Bob, known as Holevo bound, and $\Delta(n)$ is the offset term accounting for privacy amplification in finite-size regime. The worst-case scenario is adopted to estimate Eve's Holevo bound by building corresponding maximum-likelihood estimators for channel parameters, which are bounded by 6.5 standard deviations.
One-time SNU calibration is used to simplify the calibration process and achieve a real-time SNU result, which can effectively avoid the mis-estimation of the SNU with fluctuations \cite{Zhang_PhysRevApplied_2020}. Here, SNU is redefined as the sum of shot noise and electronic noise, while only the detection efficiency is trusted modeled.
The overall detection efficiency of the receiver is 0.2271, contributed by the responsibility (0.8 A/W), coupling loss (4 dB) as well as the extra loss from PC and the connectors (0.5 dB). The other losses introduced by the VOAs before the PDs, and the electronic noise, which is transferred to channel loss in the one-time SNU calibration scenario, are considered as the untrusted loss for worst-case estimation, which is 5 dB in total.
For the enhancement of the signal-to-noise ratio, the modulation variance is set as 8 SNU and the reconciliation efficiency is $95.6 \ \%$ \cite{yang2023information}.
The repetition frequency is 1 GHz, and the overhead is $50 \ \%$ due to the frame structure, SNU calibration and parameter estimation.
The key parameters affecting the system performance, the excess noise $\varepsilon$, is 0.055 SNU in this experiment, and the maximum length of the data is $10^9$.

The asymptotic and finite-size secret key rates of the system are shown in Fig. \ref{fig:SKR}, where the transmission distance reaches 28.6 km with an asymptotic secret key rate of 1.38 Mbps. When considering the finite-size effect with the data length of $10^{10}$ and $10^{9}$, the secret key rate can be fixed as 0.54 and 0.24 Mbps.
Compared with the existing results \cite{pietri2023cv}, the presented experiment has increased the transmission distance by 4 times. This advancement is brought by the suppression of the excess noise at the receiver site, contributed by the high-performance low-noise integrated silicon photonic receiver, the accurate real-time SNU calibration and the advanced DSP routine.
Simulations have shown the potential of extending the transmission distance to 50 km with longer data.

In conclusion, our experiment has extended the transmission distance of a chip-based CV-QKD system to 28.6 km with Mbps level asymptotic secret key rate, which can fully satisfy the demands of large-scale deployments in a CV-QKD access network, where massive high-performance  receivers with low cost and tiny size are required \cite{bian2023high,PTMPAccessNetwork,bian2023first}. The system can also support the metropolitan interconnections in short range, while for long-haul transmissions, improvements can be achieved by further optimizing the chip-based devices, including raising the overall detection efficiency by edge coupler (loss less than 1 dB) and high-responsivity PDs (responsivity up to 1.1 A/W). Predictably, combining with the chip-based laser source \cite{li2023continuous}, modulator \cite{CvExpSOICVQKD2019} and polarization controller \cite{wang2022silicon}, a fully integrated CV-QKD system on chip is not far way.
Notice that two recent works have also tested the high-performance CV-QKD system with chip-based receivers \cite{piétri2023experimental,hajomer2023continuous}, but both experimental transmission distances with stable secret key generation are around 10 km.
Our work has demonstrated the importance and feasibility of the high-performance photonic integrated receiver in a practical CV-QKD system with considerable transmission distance and secret key rate, which paves the way for the next generation CV-QKD system with high performance, compact size, and scalability.


%
%

%

\begin{acknowledgments}
    This work was supported in part by the National Key Research and Development Program of China (Grant No. 2020YFA0309704), the National Natural Science Foundation of China (Grant Nos U22A2089, 62001044, 62101516, 62171418, 62201530 and 61901425), the Sichuan Science and Technology Program (Grant Nos 2023JDRC0017, 2023YFG0143, 2022ZDZX0009 and 2021YJ0313), the Natural Science Foundation of Sichuan Province (Grant Nos 2023NSFSC1387 and 2023NSFSC0449), the Basic Research Program of China(Grant No. JCKY2021210B059), the Equipment Advance Research Field Foundation(Grant No. 315067206), the Chengdu Key Research and Development Support Program (Grant No 2021-YF09-00116-GX).
\end{acknowledgments}

\section*{DATA AVAILABILITY}
The data that support the ﬁndings of this study are available from the corresponding authors upon reasonable request.

\bibliography{aipsamp}

\providecommand{\noopsort}[1]{}\providecommand{\singleletter}[1]{#1}%
\begin{thebibliography}{40}%
\makeatletter
\providecommand \@ifxundefined [1]{%
 \@ifx{#1\undefined}
}%
\providecommand \@ifnum [1]{%
 \ifnum #1\expandafter \@firstoftwo
 \else \expandafter \@secondoftwo
 \fi
}%
\providecommand \@ifx [1]{%
 \ifx #1\expandafter \@firstoftwo
 \else \expandafter \@secondoftwo
 \fi
}%
\providecommand \natexlab [1]{#1}%
\providecommand \enquote  [1]{``#1''}%
\providecommand \bibnamefont  [1]{#1}%
\providecommand \bibfnamefont [1]{#1}%
\providecommand \citenamefont [1]{#1}%
\providecommand \href@noop [0]{\@secondoftwo}%
\providecommand \href [0]{\begingroup \@sanitize@url \@href}%
\providecommand \@href[1]{\@@startlink{#1}\@@href}%
\providecommand \@@href[1]{\endgroup#1\@@endlink}%
\providecommand \@sanitize@url [0]{\catcode `\\12\catcode `\$12\catcode `\&12\catcode `\#12\catcode `\^12\catcode `\_12\catcode `\%12\relax}%
\providecommand \@@startlink[1]{}%
\providecommand \@@endlink[0]{}%
\providecommand \url  [0]{\begingroup\@sanitize@url \@url }%
\providecommand \@url [1]{\endgroup\@href {#1}{\urlprefix }}%
\providecommand \urlprefix  [0]{URL }%
\providecommand \Eprint [0]{\href }%
\providecommand \doibase [0]{http://dx.doi.org/}%
\providecommand \selectlanguage [0]{\@gobble}%
\providecommand \bibinfo  [0]{\@secondoftwo}%
\providecommand \bibfield  [0]{\@secondoftwo}%
\providecommand \translation [1]{[#1]}%
\providecommand \BibitemOpen [0]{}%
\providecommand \bibitemStop [0]{}%
\providecommand \bibitemNoStop [0]{.\EOS\space}%
\providecommand \EOS [0]{\spacefactor3000\relax}%
\providecommand \BibitemShut  [1]{\csname bibitem#1\endcsname}%
\let\auto@bib@innerbib\@empty
\bibitem [{\citenamefont {Bennet}\ and\ \citenamefont {Brassard}(1984)}]{bennet1984quantum}%
  \BibitemOpen
  \bibfield  {author} {\bibinfo {author} {\bibfnamefont {C.}~\bibnamefont {Bennet}}\ and\ \bibinfo {author} {\bibfnamefont {G.}~\bibnamefont {Brassard}},\ }\bibfield  {title} {\enquote {\bibinfo {title} {Quantum cryptography: Public key distribution and coin tossing},}\ }in\ \href@noop {} {\emph {\bibinfo {booktitle} {Proceedings of IEEE International Conference on Computers, Systems, and Signal Processing, Bangalore, India}}}\ (\bibinfo {year} {1984})\ pp.\ \bibinfo {pages} {175--179}\BibitemShut {NoStop}%
\bibitem [{\citenamefont {Pirandola}\ \emph {et~al.}(2020)\citenamefont {Pirandola}, \citenamefont {Andersen}, \citenamefont {Banchi} \emph {et~al.}}]{AdvInQC}%
  \BibitemOpen
  \bibfield  {author} {\bibinfo {author} {\bibfnamefont {S.}~\bibnamefont {Pirandola}}, \bibinfo {author} {\bibfnamefont {U.~L.}\ \bibnamefont {Andersen}}, \bibinfo {author} {\bibfnamefont {L.}~\bibnamefont {Banchi}},  \emph {et~al.},\ }\bibfield  {title} {\enquote {\bibinfo {title} {Advances in quantum cryptography},}\ }\href {\doibase 10.1364/AOP.361502} {\bibfield  {journal} {\bibinfo  {journal} {Adv. Opt. Photon.}\ }\textbf {\bibinfo {volume} {12}},\ \bibinfo {pages} {1012--1236} (\bibinfo {year} {2020})}\BibitemShut {NoStop}%
\bibitem [{\citenamefont {Xu}\ \emph {et~al.}(2020)\citenamefont {Xu}, \citenamefont {Ma}, \citenamefont {Zhang} \emph {et~al.}}]{PTPQKDRMV2020}%
  \BibitemOpen
  \bibfield  {author} {\bibinfo {author} {\bibfnamefont {F.}~\bibnamefont {Xu}}, \bibinfo {author} {\bibfnamefont {X.}~\bibnamefont {Ma}}, \bibinfo {author} {\bibfnamefont {Q.}~\bibnamefont {Zhang}},  \emph {et~al.},\ }\bibfield  {title} {\enquote {\bibinfo {title} {Secure quantum key distribution with realistic devices},}\ }\href {\doibase 10.1103/RevModPhys.92.025002} {\bibfield  {journal} {\bibinfo  {journal} {Rev. Mod. Phys.}\ }\textbf {\bibinfo {volume} {92}},\ \bibinfo {pages} {025002} (\bibinfo {year} {2020})}\BibitemShut {NoStop}%
\bibitem [{\citenamefont {Grosshans}\ \emph {et~al.}(2003)\citenamefont {Grosshans}, \citenamefont {Wenger}, \citenamefont {Tualle-Brouri} \emph {et~al.}}]{GG02Nature}%
  \BibitemOpen
  \bibfield  {author} {\bibinfo {author} {\bibfnamefont {F.}~\bibnamefont {Grosshans}}, \bibinfo {author} {\bibfnamefont {J.}~\bibnamefont {Wenger}}, \bibinfo {author} {\bibfnamefont {R.}~\bibnamefont {Tualle-Brouri}},  \emph {et~al.},\ }\bibfield  {title} {\enquote {\bibinfo {title} {Quantum key distribution using gaussian-modulated coherent states},}\ }\href {\doibase https://doi.org/10.1038/nature01289} {\bibfield  {journal} {\bibinfo  {journal} {Nature}\ }\textbf {\bibinfo {volume} {421}},\ \bibinfo {pages} {238--241} (\bibinfo {year} {2003})}\BibitemShut {NoStop}%
\bibitem [{\citenamefont {Weedbrook}\ \emph {et~al.}(2011)\citenamefont {Weedbrook}, \citenamefont {Pirandola}, \citenamefont {Garcia-Patron} \emph {et~al.}}]{GaussianQuantumInformation}%
  \BibitemOpen
  \bibfield  {author} {\bibinfo {author} {\bibfnamefont {C.}~\bibnamefont {Weedbrook}}, \bibinfo {author} {\bibfnamefont {S.}~\bibnamefont {Pirandola}}, \bibinfo {author} {\bibfnamefont {R.}~\bibnamefont {Garcia-Patron}},  \emph {et~al.},\ }\bibfield  {title} {\enquote {\bibinfo {title} {Gaussian quantum information},}\ }\href {\doibase 10.1103/RevModPhys.84.621} {\bibfield  {journal} {\bibinfo  {journal} {Rev. Mod. Phys.}\ }\textbf {\bibinfo {volume} {84}},\ \bibinfo {pages} {621} (\bibinfo {year} {2011})}\BibitemShut {NoStop}%
\bibitem [{\citenamefont {Lam}\ and\ \citenamefont {Ralph}(2013)}]{lam2013continuous}%
  \BibitemOpen
  \bibfield  {author} {\bibinfo {author} {\bibfnamefont {P.~K.}\ \bibnamefont {Lam}}\ and\ \bibinfo {author} {\bibfnamefont {T.~C.}\ \bibnamefont {Ralph}},\ }\bibfield  {title} {\enquote {\bibinfo {title} {Continuous improvement},}\ }\href {\doibase https://doi.org/10.1038/nphoton.2013.104} {\bibfield  {journal} {\bibinfo  {journal} {Nat. Photonics}\ }\textbf {\bibinfo {volume} {7}},\ \bibinfo {pages} {350--352} (\bibinfo {year} {2013})}\BibitemShut {NoStop}%
\bibitem [{\citenamefont {Grosshans}\ and\ \citenamefont {Grangier}(2002)}]{GG02PRL}%
  \BibitemOpen
  \bibfield  {author} {\bibinfo {author} {\bibfnamefont {F.}~\bibnamefont {Grosshans}}\ and\ \bibinfo {author} {\bibfnamefont {P.}~\bibnamefont {Grangier}},\ }\bibfield  {title} {\enquote {\bibinfo {title} {Continuous variable quantum cryptography using coherent states},}\ }\href {\doibase 10.1103/PhysRevLett.88.057902} {\bibfield  {journal} {\bibinfo  {journal} {Phys. Rev. Lett.}\ }\textbf {\bibinfo {volume} {88}},\ \bibinfo {pages} {057902} (\bibinfo {year} {2002})}\BibitemShut {NoStop}%
\bibitem [{\citenamefont {Weedbrook}\ \emph {et~al.}(2004)\citenamefont {Weedbrook}, \citenamefont {Lance}, \citenamefont {Bowen} \emph {et~al.}}]{NSPRL}%
  \BibitemOpen
  \bibfield  {author} {\bibinfo {author} {\bibfnamefont {C.}~\bibnamefont {Weedbrook}}, \bibinfo {author} {\bibfnamefont {A.~M.}\ \bibnamefont {Lance}}, \bibinfo {author} {\bibfnamefont {W.~P.}\ \bibnamefont {Bowen}},  \emph {et~al.},\ }\bibfield  {title} {\enquote {\bibinfo {title} {Quantum cryptography without switching},}\ }\href {\doibase 10.1103/PhysRevLett.93.170504} {\bibfield  {journal} {\bibinfo  {journal} {Phys. Rev. Lett.}\ }\textbf {\bibinfo {volume} {93}},\ \bibinfo {pages} {170504} (\bibinfo {year} {2004})}\BibitemShut {NoStop}%
\bibitem [{\citenamefont {Jouguet}\ \emph {et~al.}(2012)\citenamefont {Jouguet}, \citenamefont {Kunz-Jacques}, \citenamefont {Leverrier} \emph {et~al.}}]{CvExp80km2012}%
  \BibitemOpen
  \bibfield  {author} {\bibinfo {author} {\bibfnamefont {P.}~\bibnamefont {Jouguet}}, \bibinfo {author} {\bibfnamefont {S.}~\bibnamefont {Kunz-Jacques}}, \bibinfo {author} {\bibfnamefont {A.}~\bibnamefont {Leverrier}},  \emph {et~al.},\ }\bibfield  {title} {\enquote {\bibinfo {title} {Experimental demonstration of long-distance continuous-variable quantum key distribution},}\ }\href {\doibase 10.1038/NPHOTON.2013.63} {\bibfield  {journal} {\bibinfo  {journal} {Nat. Photonics}\ }\textbf {\bibinfo {volume} {7}},\ \bibinfo {pages} {378--381} (\bibinfo {year} {2012})}\BibitemShut {NoStop}%
\bibitem [{\citenamefont {Pirandola}\ \emph {et~al.}(2015)\citenamefont {Pirandola}, \citenamefont {Ottaviani}, \citenamefont {Spedalieri} \emph {et~al.}}]{CVMDIYork}%
  \BibitemOpen
  \bibfield  {author} {\bibinfo {author} {\bibfnamefont {S.}~\bibnamefont {Pirandola}}, \bibinfo {author} {\bibfnamefont {C.}~\bibnamefont {Ottaviani}}, \bibinfo {author} {\bibfnamefont {G.}~\bibnamefont {Spedalieri}},  \emph {et~al.},\ }\bibfield  {title} {\enquote {\bibinfo {title} {High-rate measurement-device-independent quantum cryptography},}\ }\href {\doibase 10.1038/nphoton.2015.83} {\bibfield  {journal} {\bibinfo  {journal} {Nat. Photonics}\ }\textbf {\bibinfo {volume} {9}},\ \bibinfo {pages} {397--402} (\bibinfo {year} {2015})}\BibitemShut {NoStop}%
\bibitem [{\citenamefont {Zhang}\ \emph {et~al.}(2019{\natexlab{a}})\citenamefont {Zhang}, \citenamefont {Li}, \citenamefont {Chen} \emph {et~al.}}]{CVQKD50km}%
  \BibitemOpen
  \bibfield  {author} {\bibinfo {author} {\bibfnamefont {Y.}~\bibnamefont {Zhang}}, \bibinfo {author} {\bibfnamefont {Z.}~\bibnamefont {Li}}, \bibinfo {author} {\bibfnamefont {Z.}~\bibnamefont {Chen}},  \emph {et~al.},\ }\bibfield  {title} {\enquote {\bibinfo {title} {Continuous-variable {QKD} over 50 km commercial fiber},}\ }\href {\doibase 10.1088/2058-9565/ab19d1} {\bibfield  {journal} {\bibinfo  {journal} {Quantum Sci. Technol.}\ }\textbf {\bibinfo {volume} {4}},\ \bibinfo {pages} {035006} (\bibinfo {year} {2019}{\natexlab{a}})}\BibitemShut {NoStop}%
\bibitem [{\citenamefont {Zhang}\ \emph {et~al.}(2020{\natexlab{a}})\citenamefont {Zhang}, \citenamefont {Chen}, \citenamefont {Pirandola} \emph {et~al.}}]{CvExp202kmPRL}%
  \BibitemOpen
  \bibfield  {author} {\bibinfo {author} {\bibfnamefont {Y.}~\bibnamefont {Zhang}}, \bibinfo {author} {\bibfnamefont {Z.}~\bibnamefont {Chen}}, \bibinfo {author} {\bibfnamefont {S.}~\bibnamefont {Pirandola}},  \emph {et~al.},\ }\bibfield  {title} {\enquote {\bibinfo {title} {Long-distance continuous-variable quantum key distribution over 202.81 km of fiber},}\ }\href {\doibase 10.1103/PhysRevLett.125.010502} {\bibfield  {journal} {\bibinfo  {journal} {Phys. Rev. Lett.}\ }\textbf {\bibinfo {volume} {125}},\ \bibinfo {pages} {010502} (\bibinfo {year} {2020}{\natexlab{a}})}\BibitemShut {NoStop}%
\bibitem [{\citenamefont {Jain}\ \emph {et~al.}(2022)\citenamefont {Jain}, \citenamefont {Chin}, \citenamefont {Mani} \emph {et~al.}}]{CVNC2022}%
  \BibitemOpen
  \bibfield  {author} {\bibinfo {author} {\bibfnamefont {N.}~\bibnamefont {Jain}}, \bibinfo {author} {\bibfnamefont {H.-M.}\ \bibnamefont {Chin}}, \bibinfo {author} {\bibfnamefont {H.}~\bibnamefont {Mani}},  \emph {et~al.},\ }\bibfield  {title} {\enquote {\bibinfo {title} {Practical continuous-variable quantum key distribution with composable security},}\ }\href {\doibase 10.1038/s41467-022-32161-y} {\bibfield  {journal} {\bibinfo  {journal} {Nat. Commun.}\ }\textbf {\bibinfo {volume} {13}},\ \bibinfo {pages} {4740} (\bibinfo {year} {2022})}\BibitemShut {NoStop}%
\bibitem [{\citenamefont {Hajomer}\ \emph {et~al.}(2024)\citenamefont {Hajomer}, \citenamefont {Derkach}, \citenamefont {Jain} \emph {et~al.}}]{hajomer2024long}%
  \BibitemOpen
  \bibfield  {author} {\bibinfo {author} {\bibfnamefont {A.~A.}\ \bibnamefont {Hajomer}}, \bibinfo {author} {\bibfnamefont {I.}~\bibnamefont {Derkach}}, \bibinfo {author} {\bibfnamefont {N.}~\bibnamefont {Jain}},  \emph {et~al.},\ }\bibfield  {title} {\enquote {\bibinfo {title} {Long-distance continuous-variable quantum key distribution over 100-km fiber with local local oscillator},}\ }\href@noop {} {\bibfield  {journal} {\bibinfo  {journal} {Sci. Adv.}\ }\textbf {\bibinfo {volume} {10}},\ \bibinfo {pages} {eadi9474} (\bibinfo {year} {2024})}\BibitemShut {NoStop}%
\bibitem [{\citenamefont {Ghorai}\ \emph {et~al.}(2019)\citenamefont {Ghorai}, \citenamefont {Grangier}, \citenamefont {Diamanti} \emph {et~al.}}]{DMCVLeverrier}%
  \BibitemOpen
  \bibfield  {author} {\bibinfo {author} {\bibfnamefont {S.}~\bibnamefont {Ghorai}}, \bibinfo {author} {\bibfnamefont {P.}~\bibnamefont {Grangier}}, \bibinfo {author} {\bibfnamefont {E.}~\bibnamefont {Diamanti}},  \emph {et~al.},\ }\bibfield  {title} {\enquote {\bibinfo {title} {Asymptotic security of continuous-variable quantum key distribution with a discrete modulation},}\ }\href@noop {} {\bibfield  {journal} {\bibinfo  {journal} {Phys. Rev. X}\ }\textbf {\bibinfo {volume} {9}} (\bibinfo {year} {2019})}\BibitemShut {NoStop}%
\bibitem [{\citenamefont {Lin}, \citenamefont {Upadhyaya},\ and\ \citenamefont {Lütkenhaus}(2019)}]{DMCVLinjie}%
  \BibitemOpen
  \bibfield  {author} {\bibinfo {author} {\bibfnamefont {J.}~\bibnamefont {Lin}}, \bibinfo {author} {\bibfnamefont {T.}~\bibnamefont {Upadhyaya}}, \ and\ \bibinfo {author} {\bibfnamefont {N.}~\bibnamefont {Lütkenhaus}},\ }\bibfield  {title} {\enquote {\bibinfo {title} {Asymptotic security analysis of discrete-modulated continuous-variable quantum key distribution},}\ }\href@noop {} {\bibfield  {journal} {\bibinfo  {journal} {Phys. Rev. X}\ }\textbf {\bibinfo {volume} {9}} (\bibinfo {year} {2019})}\BibitemShut {NoStop}%
\bibitem [{\citenamefont {Wang}\ \emph {et~al.}(2022{\natexlab{a}})\citenamefont {Wang}, \citenamefont {Li}, \citenamefont {Pi} \emph {et~al.}}]{SubGbps}%
  \BibitemOpen
  \bibfield  {author} {\bibinfo {author} {\bibfnamefont {H.}~\bibnamefont {Wang}}, \bibinfo {author} {\bibfnamefont {Y.}~\bibnamefont {Li}}, \bibinfo {author} {\bibfnamefont {Y.}~\bibnamefont {Pi}},  \emph {et~al.},\ }\bibfield  {title} {\enquote {\bibinfo {title} {Sub-gbps key rate four-state continuous-variable quantum key distribution within metropolitan area},}\ }\href {\doibase 10.1038/s42005-022-00941-z} {\bibfield  {journal} {\bibinfo  {journal} {Commun. Phys.}\ }\textbf {\bibinfo {volume} {5}},\ \bibinfo {pages} {162} (\bibinfo {year} {2022}{\natexlab{a}})}\BibitemShut {NoStop}%
\bibitem [{\citenamefont {Pi}\ \emph {et~al.}(2023)\citenamefont {Pi}, \citenamefont {Wang}, \citenamefont {Pan} \emph {et~al.}}]{Pi2023SubMbps}%
  \BibitemOpen
  \bibfield  {author} {\bibinfo {author} {\bibfnamefont {Y.}~\bibnamefont {Pi}}, \bibinfo {author} {\bibfnamefont {H.}~\bibnamefont {Wang}}, \bibinfo {author} {\bibfnamefont {Y.}~\bibnamefont {Pan}},  \emph {et~al.},\ }\bibfield  {title} {\enquote {\bibinfo {title} {Sub-mbps key-rate continuous-variable quantum key distribution with local local oscillator over 100-km fiber},}\ }\href {\doibase 10.1364/OL.485913} {\bibfield  {journal} {\bibinfo  {journal} {Opt. Lett.}\ }\textbf {\bibinfo {volume} {48}},\ \bibinfo {pages} {1766--1769} (\bibinfo {year} {2023})}\BibitemShut {NoStop}%
\bibitem [{\citenamefont {Zhang}\ \emph {et~al.}(2023)\citenamefont {Zhang}, \citenamefont {Bian}, \citenamefont {Li} \emph {et~al.}}]{CVReV2023}%
  \BibitemOpen
  \bibfield  {author} {\bibinfo {author} {\bibfnamefont {Y.}~\bibnamefont {Zhang}}, \bibinfo {author} {\bibfnamefont {Y.}~\bibnamefont {Bian}}, \bibinfo {author} {\bibfnamefont {Z.}~\bibnamefont {Li}},  \emph {et~al.},\ }\bibfield  {title} {\enquote {\bibinfo {title} {Continuous-variable quantum key distribution system: A review and perspective},}\ }\href@noop {} {\bibfield  {journal} {\bibinfo  {journal} {arXiv:2310.04831}\ } (\bibinfo {year} {2023})}\BibitemShut {NoStop}%
\bibitem [{\citenamefont {Wang}\ \emph {et~al.}(2020)\citenamefont {Wang}, \citenamefont {Sciarrino}, \citenamefont {Laing} \emph {et~al.}}]{wang2020integrated}%
  \BibitemOpen
  \bibfield  {author} {\bibinfo {author} {\bibfnamefont {J.}~\bibnamefont {Wang}}, \bibinfo {author} {\bibfnamefont {F.}~\bibnamefont {Sciarrino}}, \bibinfo {author} {\bibfnamefont {A.}~\bibnamefont {Laing}},  \emph {et~al.},\ }\bibfield  {title} {\enquote {\bibinfo {title} {Integrated photonic quantum technologies},}\ }\href@noop {} {\bibfield  {journal} {\bibinfo  {journal} {Nat. Photonics}\ }\textbf {\bibinfo {volume} {14}},\ \bibinfo {pages} {273--284} (\bibinfo {year} {2020})}\BibitemShut {NoStop}%
\bibitem [{\citenamefont {Luo}\ \emph {et~al.}(2023)\citenamefont {Luo}, \citenamefont {Cao}, \citenamefont {Shi} \emph {et~al.}}]{luo2023recent}%
  \BibitemOpen
  \bibfield  {author} {\bibinfo {author} {\bibfnamefont {W.}~\bibnamefont {Luo}}, \bibinfo {author} {\bibfnamefont {L.}~\bibnamefont {Cao}}, \bibinfo {author} {\bibfnamefont {Y.}~\bibnamefont {Shi}},  \emph {et~al.},\ }\bibfield  {title} {\enquote {\bibinfo {title} {Recent progress in quantum photonic chips for quantum communication and internet},}\ }\href@noop {} {\bibfield  {journal} {\bibinfo  {journal} {Light Sci. Appl.}\ }\textbf {\bibinfo {volume} {12}},\ \bibinfo {pages} {175} (\bibinfo {year} {2023})}\BibitemShut {NoStop}%
\bibitem [{\citenamefont {Soref}(2006)}]{soref2006past}%
  \BibitemOpen
  \bibfield  {author} {\bibinfo {author} {\bibfnamefont {R.}~\bibnamefont {Soref}},\ }\bibfield  {title} {\enquote {\bibinfo {title} {The past, present, and future of silicon photonics},}\ }\href@noop {} {\bibfield  {journal} {\bibinfo  {journal} {IEEE J. Sel. Top. Quantum Electron.}\ }\textbf {\bibinfo {volume} {12}},\ \bibinfo {pages} {1678--1687} (\bibinfo {year} {2006})}\BibitemShut {NoStop}%
\bibitem [{\citenamefont {Lim}\ \emph {et~al.}(2013)\citenamefont {Lim}, \citenamefont {Song}, \citenamefont {Fang} \emph {et~al.}}]{lim2013review}%
  \BibitemOpen
  \bibfield  {author} {\bibinfo {author} {\bibfnamefont {A.~E.-J.}\ \bibnamefont {Lim}}, \bibinfo {author} {\bibfnamefont {J.}~\bibnamefont {Song}}, \bibinfo {author} {\bibfnamefont {Q.}~\bibnamefont {Fang}},  \emph {et~al.},\ }\bibfield  {title} {\enquote {\bibinfo {title} {Review of silicon photonics foundry efforts},}\ }\href@noop {} {\bibfield  {journal} {\bibinfo  {journal} {IEEE J. Sel. Top. Quantum Electron.}\ }\textbf {\bibinfo {volume} {20}},\ \bibinfo {pages} {405--416} (\bibinfo {year} {2013})}\BibitemShut {NoStop}%
\bibitem [{\citenamefont {Siew}\ \emph {et~al.}(2021)\citenamefont {Siew}, \citenamefont {Li}, \citenamefont {Gao} \emph {et~al.}}]{siew2021review}%
  \BibitemOpen
  \bibfield  {author} {\bibinfo {author} {\bibfnamefont {S.~Y.}\ \bibnamefont {Siew}}, \bibinfo {author} {\bibfnamefont {B.}~\bibnamefont {Li}}, \bibinfo {author} {\bibfnamefont {F.}~\bibnamefont {Gao}},  \emph {et~al.},\ }\bibfield  {title} {\enquote {\bibinfo {title} {Review of silicon photonics technology and platform development},}\ }\href@noop {} {\bibfield  {journal} {\bibinfo  {journal} {J. Lightwave Technol.}\ }\textbf {\bibinfo {volume} {39}},\ \bibinfo {pages} {4374--4389} (\bibinfo {year} {2021})}\BibitemShut {NoStop}%
\bibitem [{\citenamefont {Bruynsteen}\ \emph {et~al.}(2021)\citenamefont {Bruynsteen}, \citenamefont {Vanhoecke}, \citenamefont {Bauwelinck} \emph {et~al.}}]{bruynsteen2021integrated}%
  \BibitemOpen
  \bibfield  {author} {\bibinfo {author} {\bibfnamefont {C.}~\bibnamefont {Bruynsteen}}, \bibinfo {author} {\bibfnamefont {M.}~\bibnamefont {Vanhoecke}}, \bibinfo {author} {\bibfnamefont {J.}~\bibnamefont {Bauwelinck}},  \emph {et~al.},\ }\bibfield  {title} {\enquote {\bibinfo {title} {Integrated balanced homodyne photonic--electronic detector for beyond 20 ghz shot-noise-limited measurements},}\ }\href@noop {} {\bibfield  {journal} {\bibinfo  {journal} {Optica}\ }\textbf {\bibinfo {volume} {8}},\ \bibinfo {pages} {1146--1152} (\bibinfo {year} {2021})}\BibitemShut {NoStop}%
\bibitem [{\citenamefont {Tasker}\ \emph {et~al.}(2021)\citenamefont {Tasker}, \citenamefont {Frazer}, \citenamefont {Ferranti} \emph {et~al.}}]{tasker2021silicon}%
  \BibitemOpen
  \bibfield  {author} {\bibinfo {author} {\bibfnamefont {J.~F.}\ \bibnamefont {Tasker}}, \bibinfo {author} {\bibfnamefont {J.}~\bibnamefont {Frazer}}, \bibinfo {author} {\bibfnamefont {G.}~\bibnamefont {Ferranti}},  \emph {et~al.},\ }\bibfield  {title} {\enquote {\bibinfo {title} {Silicon photonics interfaced with integrated electronics for 9 ghz measurement of squeezed light},}\ }\href@noop {} {\bibfield  {journal} {\bibinfo  {journal} {Nat. Photonics}\ }\textbf {\bibinfo {volume} {15}},\ \bibinfo {pages} {11--15} (\bibinfo {year} {2021})}\BibitemShut {NoStop}%
\bibitem [{\citenamefont {Jia}\ \emph {et~al.}(2023)\citenamefont {Jia}, \citenamefont {Wang}, \citenamefont {Hu} \emph {et~al.}}]{jia2023silicon}%
  \BibitemOpen
  \bibfield  {author} {\bibinfo {author} {\bibfnamefont {Y.}~\bibnamefont {Jia}}, \bibinfo {author} {\bibfnamefont {X.}~\bibnamefont {Wang}}, \bibinfo {author} {\bibfnamefont {X.}~\bibnamefont {Hu}},  \emph {et~al.},\ }\bibfield  {title} {\enquote {\bibinfo {title} {Silicon photonics-integrated time-domain balanced homodyne detector for quantum tomography and quantum key distribution},}\ }\href@noop {} {\bibfield  {journal} {\bibinfo  {journal} {New J. Phys.}\ }\textbf {\bibinfo {volume} {25}},\ \bibinfo {pages} {103030} (\bibinfo {year} {2023})}\BibitemShut {NoStop}%
\bibitem [{\citenamefont {Zhang}\ \emph {et~al.}(2019{\natexlab{b}})\citenamefont {Zhang}, \citenamefont {Haw}, \citenamefont {Cai} \emph {et~al.}}]{CvExpSOICVQKD2019}%
  \BibitemOpen
  \bibfield  {author} {\bibinfo {author} {\bibfnamefont {G.}~\bibnamefont {Zhang}}, \bibinfo {author} {\bibfnamefont {J.~Y.}\ \bibnamefont {Haw}}, \bibinfo {author} {\bibfnamefont {H.}~\bibnamefont {Cai}},  \emph {et~al.},\ }\bibfield  {title} {\enquote {\bibinfo {title} {An integrated silicon photonic chip platform for continuous-variable quantum key distribution},}\ }\href {\doibase 10.1038/s41566-019-0504-5} {\bibfield  {journal} {\bibinfo  {journal} {Nat. Photonics}\ }\textbf {\bibinfo {volume} {13}},\ \bibinfo {pages} {839--842} (\bibinfo {year} {2019}{\natexlab{b}})}\BibitemShut {NoStop}%
\bibitem [{\citenamefont {Pi{\'e}tri}\ \emph {et~al.}(2023)\citenamefont {Pi{\'e}tri}, \citenamefont {Vidarte}, \citenamefont {Schiavon} \emph {et~al.}}]{pietri2023cv}%
  \BibitemOpen
  \bibfield  {author} {\bibinfo {author} {\bibfnamefont {Y.}~\bibnamefont {Pi{\'e}tri}}, \bibinfo {author} {\bibfnamefont {L.~T.}\ \bibnamefont {Vidarte}}, \bibinfo {author} {\bibfnamefont {M.}~\bibnamefont {Schiavon}},  \emph {et~al.},\ }\bibfield  {title} {\enquote {\bibinfo {title} {Cv-qkd receiver platform based on a silicon photonic integrated circuit},}\ }in\ \href@noop {} {\emph {\bibinfo {booktitle} {2023 Optical Fiber Communications Conference and Exhibition (OFC)}}}\ (\bibinfo {organization} {IEEE},\ \bibinfo {year} {2023})\ pp.\ \bibinfo {pages} {1--3}\BibitemShut {NoStop}%
\bibitem [{SIT(2024)}]{SITRI}%
  \BibitemOpen
  \href@noop {} {\enquote {\bibinfo {title} {The homepage of sitri},}\ } (\bibinfo {year} {2024}),\ \bibinfo {note} {\url{https://www.sitri.com/about-sitri/introduction/}, Last accessed on 2024-02-08}\BibitemShut {NoStop}%
\bibitem [{\citenamefont {Pirandola}\ \emph {et~al.}(2017)\citenamefont {Pirandola}, \citenamefont {Laurenza}, \citenamefont {Ottaviani} \emph {et~al.}}]{pirandola2017fundamental}%
  \BibitemOpen
  \bibfield  {author} {\bibinfo {author} {\bibfnamefont {S.}~\bibnamefont {Pirandola}}, \bibinfo {author} {\bibfnamefont {R.}~\bibnamefont {Laurenza}}, \bibinfo {author} {\bibfnamefont {C.}~\bibnamefont {Ottaviani}},  \emph {et~al.},\ }\bibfield  {title} {\enquote {\bibinfo {title} {Fundamental limits of repeaterless quantum communications},}\ }\href@noop {} {\bibfield  {journal} {\bibinfo  {journal} {Nat. commun.}\ }\textbf {\bibinfo {volume} {8}},\ \bibinfo {pages} {15043} (\bibinfo {year} {2017})}\BibitemShut {NoStop}%
\bibitem [{\citenamefont {Zhang}\ \emph {et~al.}(2020{\natexlab{b}})\citenamefont {Zhang}, \citenamefont {Huang}, \citenamefont {Chen} \emph {et~al.}}]{Zhang_PhysRevApplied_2020}%
  \BibitemOpen
  \bibfield  {author} {\bibinfo {author} {\bibfnamefont {Y.}~\bibnamefont {Zhang}}, \bibinfo {author} {\bibfnamefont {Y.}~\bibnamefont {Huang}}, \bibinfo {author} {\bibfnamefont {Z.}~\bibnamefont {Chen}},  \emph {et~al.},\ }\bibfield  {title} {\enquote {\bibinfo {title} {One-time shot-noise unit calibration method for continuous-variable quantum key distribution},}\ }\href@noop {} {\bibfield  {journal} {\bibinfo  {journal} {Phys. Rev. Appl.}\ }\textbf {\bibinfo {volume} {13}},\ \bibinfo {pages} {024058} (\bibinfo {year} {2020}{\natexlab{b}})}\BibitemShut {NoStop}%
\bibitem [{\citenamefont {Yang}\ \emph {et~al.}(2023)\citenamefont {Yang}, \citenamefont {Yan}, \citenamefont {Yang} \emph {et~al.}}]{yang2023information}%
  \BibitemOpen
  \bibfield  {author} {\bibinfo {author} {\bibfnamefont {S.}~\bibnamefont {Yang}}, \bibinfo {author} {\bibfnamefont {Z.}~\bibnamefont {Yan}}, \bibinfo {author} {\bibfnamefont {H.}~\bibnamefont {Yang}},  \emph {et~al.},\ }\bibfield  {title} {\enquote {\bibinfo {title} {Information reconciliation of continuous-variables quantum key distribution: principles, implementations and applications},}\ }\href@noop {} {\bibfield  {journal} {\bibinfo  {journal} {EPJ Quantum Technol.}\ }\textbf {\bibinfo {volume} {10}},\ \bibinfo {pages} {40} (\bibinfo {year} {2023})}\BibitemShut {NoStop}%
\bibitem [{\citenamefont {Bian}\ \emph {et~al.}(2023{\natexlab{a}})\citenamefont {Bian}, \citenamefont {Zhang}, \citenamefont {Zhou} \emph {et~al.}}]{bian2023high}%
  \BibitemOpen
  \bibfield  {author} {\bibinfo {author} {\bibfnamefont {Y.}~\bibnamefont {Bian}}, \bibinfo {author} {\bibfnamefont {Y.-C.}\ \bibnamefont {Zhang}}, \bibinfo {author} {\bibfnamefont {C.}~\bibnamefont {Zhou}},  \emph {et~al.},\ }\href@noop {} {\enquote {\bibinfo {title} {High-rate point-to-multipoint quantum key distribution using coherent states},}\ } (\bibinfo {year} {2023}{\natexlab{a}}),\ \Eprint {http://arxiv.org/abs/2302.02391} {arXiv:2302.02391 [quant-ph]} \BibitemShut {NoStop}%
\bibitem [{\citenamefont {Pan}\ \emph {et~al.}(2023)\citenamefont {Pan}, \citenamefont {Bian}, \citenamefont {Wang} \emph {et~al.}}]{PTMPAccessNetwork}%
  \BibitemOpen
  \bibfield  {author} {\bibinfo {author} {\bibfnamefont {Y.}~\bibnamefont {Pan}}, \bibinfo {author} {\bibfnamefont {Y.}~\bibnamefont {Bian}}, \bibinfo {author} {\bibfnamefont {H.}~\bibnamefont {Wang}},  \emph {et~al.},\ }\bibfield  {title} {\enquote {\bibinfo {title} {Experimental demonstration of 4-user quantum access network based on passive optical network},}\ }in\ \href@noop {} {\emph {\bibinfo {booktitle} {European Conference on Optical Communications (ECOC 2023)}}}\ (\bibinfo  {publisher} {IEEE},\ \bibinfo {year} {2023})\ p.\ \bibinfo {pages} {P51}\BibitemShut {NoStop}%
\bibitem [{\citenamefont {Bian}\ \emph {et~al.}(2023{\natexlab{b}})\citenamefont {Bian}, \citenamefont {Pan}, \citenamefont {Ma} \emph {et~al.}}]{bian2023first}%
  \BibitemOpen
  \bibfield  {author} {\bibinfo {author} {\bibfnamefont {Y.}~\bibnamefont {Bian}}, \bibinfo {author} {\bibfnamefont {Y.}~\bibnamefont {Pan}}, \bibinfo {author} {\bibfnamefont {L.}~\bibnamefont {Ma}},  \emph {et~al.},\ }\bibfield  {title} {\enquote {\bibinfo {title} {First demonstration of an 8-node mbps quantum access network based on passive optical distribution network facilities},}\ }in\ \href@noop {} {\emph {\bibinfo {booktitle} {Frontiers in Optics and Laser Science}}}\ (\bibinfo {organization} {Optica Publishing Group},\ \bibinfo {year} {2023})\ pp.\ \bibinfo {pages} {JTu7A--4}\BibitemShut {NoStop}%
\bibitem [{\citenamefont {Li}\ \emph {et~al.}(2023)\citenamefont {Li}, \citenamefont {Wang}, \citenamefont {Li} \emph {et~al.}}]{li2023continuous}%
  \BibitemOpen
  \bibfield  {author} {\bibinfo {author} {\bibfnamefont {L.}~\bibnamefont {Li}}, \bibinfo {author} {\bibfnamefont {T.}~\bibnamefont {Wang}}, \bibinfo {author} {\bibfnamefont {X.}~\bibnamefont {Li}},  \emph {et~al.},\ }\bibfield  {title} {\enquote {\bibinfo {title} {Continuous-variable quantum key distribution with on-chip light sources},}\ }\href@noop {} {\bibfield  {journal} {\bibinfo  {journal} {Photonics Res.}\ }\textbf {\bibinfo {volume} {11}},\ \bibinfo {pages} {504--516} (\bibinfo {year} {2023})}\BibitemShut {NoStop}%
\bibitem [{\citenamefont {Wang}\ \emph {et~al.}(2022{\natexlab{b}})\citenamefont {Wang}, \citenamefont {Jia}, \citenamefont {Guo} \emph {et~al.}}]{wang2022silicon}%
  \BibitemOpen
  \bibfield  {author} {\bibinfo {author} {\bibfnamefont {X.}~\bibnamefont {Wang}}, \bibinfo {author} {\bibfnamefont {Y.}~\bibnamefont {Jia}}, \bibinfo {author} {\bibfnamefont {X.}~\bibnamefont {Guo}},  \emph {et~al.},\ }\bibfield  {title} {\enquote {\bibinfo {title} {Silicon photonics integrated dynamic polarization controller},}\ }\href@noop {} {\bibfield  {journal} {\bibinfo  {journal} {Chinese Opt. Lett.}\ }\textbf {\bibinfo {volume} {20}},\ \bibinfo {pages} {041301} (\bibinfo {year} {2022}{\natexlab{b}})}\BibitemShut {NoStop}%
\bibitem [{\citenamefont {Piétri}\ \emph {et~al.}(2023)\citenamefont {Piétri}, \citenamefont {Vidarte}, \citenamefont {Schiavon} \emph {et~al.}}]{piétri2023experimental}%
  \BibitemOpen
  \bibfield  {author} {\bibinfo {author} {\bibfnamefont {Y.}~\bibnamefont {Piétri}}, \bibinfo {author} {\bibfnamefont {L.~T.}\ \bibnamefont {Vidarte}}, \bibinfo {author} {\bibfnamefont {M.}~\bibnamefont {Schiavon}},  \emph {et~al.},\ }\href@noop {} {\enquote {\bibinfo {title} {Experimental demonstration of continuous-variable quantum key distribution with a silicon photonics integrated receiver},}\ } (\bibinfo {year} {2023}),\ \Eprint {http://arxiv.org/abs/2311.03978} {arXiv:2311.03978 [quant-ph]} \BibitemShut {NoStop}%
\bibitem [{\citenamefont {Hajomer}\ \emph {et~al.}(2023)\citenamefont {Hajomer}, \citenamefont {Bruynsteen}, \citenamefont {Derkach} \emph {et~al.}}]{hajomer2023continuous}%
  \BibitemOpen
  \bibfield  {author} {\bibinfo {author} {\bibfnamefont {A.~A.~E.}\ \bibnamefont {Hajomer}}, \bibinfo {author} {\bibfnamefont {C.}~\bibnamefont {Bruynsteen}}, \bibinfo {author} {\bibfnamefont {I.}~\bibnamefont {Derkach}},  \emph {et~al.},\ }\href@noop {} {\enquote {\bibinfo {title} {Continuous-variable quantum key distribution at 10 gbaud using an integrated photonic-electronic receiver},}\ } (\bibinfo {year} {2023}),\ \Eprint {http://arxiv.org/abs/2305.19642} {arXiv:2305.19642 [quant-ph]} \BibitemShut {NoStop}%
\end{thebibliography}%

\end{document}